\documentclass[a4paper,11pt]{article}
\usepackage{pos}
\usepackage{lineno}
\title{The Online Observation Quality System for the ASTRI Mini-Array}
\ShortTitle{The ASTRI Mini-Array OOQS}
 
\manuallySeparateAuthors
\author*[a]{N. Parmiggiani,}
\author[a]{ A. Bulgarelli,}
\author[a]{ L. Baroncelli,}
\author[a]{ A. Addis,}
\author[a]{ V. Fioretti,}
\author[a]{ A. Di Piano,}
\author[b]{ M. Capalbi,}
\author[b]{ O. Catalano,}
\author[a]{ V. Conforti,}
\author[c,d]{ M. Fiori,}
\author[a]{ F. Gianotti,}
\author[e,f]{ S. Iovenitti,}
\author[g,h]{ F. Lucarelli,}
\author[b]{ M. C. Maccarone,}
\author[b]{ T. Mineo,}
\author[a]{ F. Russo,}
\author[b]{ P. Sangiorgi,}
\author[i]{ S. Scuderi,}
\author[l]{ G. Tosti,}
\author[a]{ M. Trifoglio}
\author[d]{ and L. Zampieri}
\author[m]{ for the ASTRI Project}

\affiliation[a]{INAF/OAS Bologna, Via P. Gobetti 93/3, I-40129 Bologna, Italy.}

\affiliation[b]{INAF/IASF Palermo, Via Ugo La Malfa 153, I-90146 Palermo, Italy.}

\affiliation[c]{Universit\`{a} degli Studi di Padova, Via VIII Febbraio 2, I-35122 Padova, Italy}

\affiliation[d]{INAF/OA Padova, Vicolo Osservatorio 5, I-35122 Padova, Italy}

\affiliation[e]{INAF/OA Brera, via Brera 28, I-20121 Milano, Italy.}

\affiliation[f]{Universit\`{a} degli Studi di Milano, Dip.to di Fisica, Via Giovanni Celoria, 16, I-20133 Milano, Italy.}

\affiliation[g]{INAF/OAR Roma, Via di Frascati 33, I-00078 Monte Porzio Catone, Roma, Italy.}

\affiliation[h]{ASI/SSDC Roma, Via del Politecnico snc, I-00133 Roma, Italy.}

\affiliation[i]{INAF/IASF Milano, Via Alfonso Corti 12, I-20133 Milano, Italy.}

\affiliation[l]{Universit\`{a} degli Studi di Perugia, Dip.to di Fisica e Geologia, Via A. Pascoli, I-06123 Perugia, Italy}

\affiliation[m]{\protect\url{http://www.astri.inaf.it}}

\emailAdd{nicolo.parmiggiani@inaf.it}

\abstract{The ASTRI Mini-Array is an international collaboration led by the Italian National Institute for Astrophysics (INAF), aiming to construct and operate an array of nine Imaging Atmospheric Cherenkov Telescopes (IACTs) to study gamma-ray sources at very high energy (TeV) and to perform stellar intensity interferometry observations. This contribution describes the design and the technologies used by the ASTRI team to implement the Online Observation Quality System (OOQS). The main objective of the OOQS is to perform data quality analyses in real-time during Cherenkov and intensity interferometry observations to provide feedback to both the Central Control System and the Operator. The OOQS performs the analysis of key data quality parameters and can generate alarms to other sub-systems for a fast reaction to solve critical conditions. The results from the data quality analyses are saved into the Quality Archive for further investigations. The Operator can visualise the OOQS results through the Operator Human Machine Interface as soon as they are produced. The main challenge addressed by the OOQS design is to perform online data quality checks on the data streams produced by nine telescopes, acquired by the Array Data Acquisition System and forwarded to the OOQS. In the current OOQS design, the Redis in-memory database manages the data throughput generated by the telescopes, and the Slurm workload scheduler executes in parallel the high number of data quality analyses.}

\FullConference{37$^{\rm{th}}$ International Cosmic Ray Conference (ICRC 2021)\\
		July 12th -- 23rd, 2021\\
		Online -- Berlin, Germany}

\begin{document}
\maketitle

\section{Introduction}\label{sec:intro}

The ASTRI Mini-Array (ASTRI MA, \cite{2016JPhCS.718e2028P}, \cite{2021ICRCAntonelli}) is an INAF project aiming to construct and operate an experiment to study gamma-ray sources emitting at very high energy in the TeV spectral band and to perform intensity interferometry observations. The ASTRI MA consists of an array of nine innovative dual-mirror Imaging Atmospheric Cherenkov Telescopes (IACTs) \citep{Krennrich_2009}. Each telescope will be equipped with the new ASTRICAM Silicon photomultiplier Cherenkov camera with its fast read-out electronics specifically designed \cite{2018SPIE10702E..37C} and managed by the Cherenkov Camera Software Supervisor \cite{2021ICRCCorpora}. Furthermore, the ASTRI-MA will perform intensity interferometry observations of a selected sample of bright sources being each telescope equipped with a Stellar Intensity Interferometry Instrument (SI$^3$). The ASTRI MA's nine telescopes will be distributed at one hundred meters of distance from each other at the Teide Astronomical Observatory, operated by the Instituto de Astrofisica de Canarias (IAC), on Mount Teide (2400 m a.s.l.) in Tenerife (Canary Islands, Spain) and INAF will operate it on the basis of a host agreement with the IAC. 

The ASTRI MA must be operated remotely, and no human presence is foreseen on-site during observations. A data centre will be installed on-site to have computing power close to the telescopes. The data acquired from the telescopes are sent to the on-site data centre for the online analysis to check the data quality and, during the observations, provide feedback both to the Operator and to the Central Control System, depending on the severity of the anomaly. 

This contribution describes the software architecture and design of the Online Observation Quality System (OOQS) for the ASTRI MA project. The OOQS aims to perform the data quality check in real-time during the observations and inform the specific ASTRI MA sub-systems if abnormal conditions are detected within the data. With this workflow, it is possible to take automated corrective actions or notify the Operator if the data acquired by the telescopes do not respect the quality standards.

Section \ref{sec:architecture} describes the software architecture of the OOQS and the design of all its software components. The software development plan and the continuous integration (CI) technologies are described in Section \ref{sec:dev_plan}. Conclusions are reported in Section \ref{sec:conclusion}.

\section{OOQS Architecture and Design}\label{sec:architecture}

The OOQS is a software system that aims to verify the online data quality during the observations performed by the ASTRI MA telescopes. The OOQS is part of the ASTRI MA Supervisory Control and Data Acquisition system (SCADA). SCADA is the system that shall manage startup, shutdown, configuration, and control all site assemblies and sub-systems to collect monitoring points, manage alarms raised by any assembly, check the health status of all systems, and acquire scientific data. Both SCADA and OOQS will be installed in the on-site data centre. The OOQS shall perform data quality checks during the Cherenkov and the intensity interferometry observations alternatively. The OOQS receives input data from the Array Data Acquisition System (ADAS), which is the ASTRI MA system designated to acquire and manage the raw data from the Cherenkov cameras and the SI$^3$ instruments. 

The OOQS system architecture comprises the following software components: OOQS Master, OOQS Manager, Cherenkov Camera Data Quality Checker (CCDQC), SI3 Data Quality Checker (SI3DQC). The first two components are implemented with the Alma Common Software (ACS, \cite{2002SPIE.4848...43C}). ACS provides a framework for the distributed development and execution of software systems. It enables the separation of functional and technical aspects so that developers can concentrate on functional elements while using the common technical features provided by ACS. The CCDQC and the SI3DQC are software components external to ACS but controlled by the OOQS Manager that can start, stop and monitor them. They are not executed simultaneously because the ASTRI MA can observe with the SI$^3$ instruments or with the Cherenkov cameras alternatively. There will be one instance of OOQS for each telescope of the array. 

\begin{figure*}[!htb]
	\centering
	  \includegraphics[width=\textwidth]{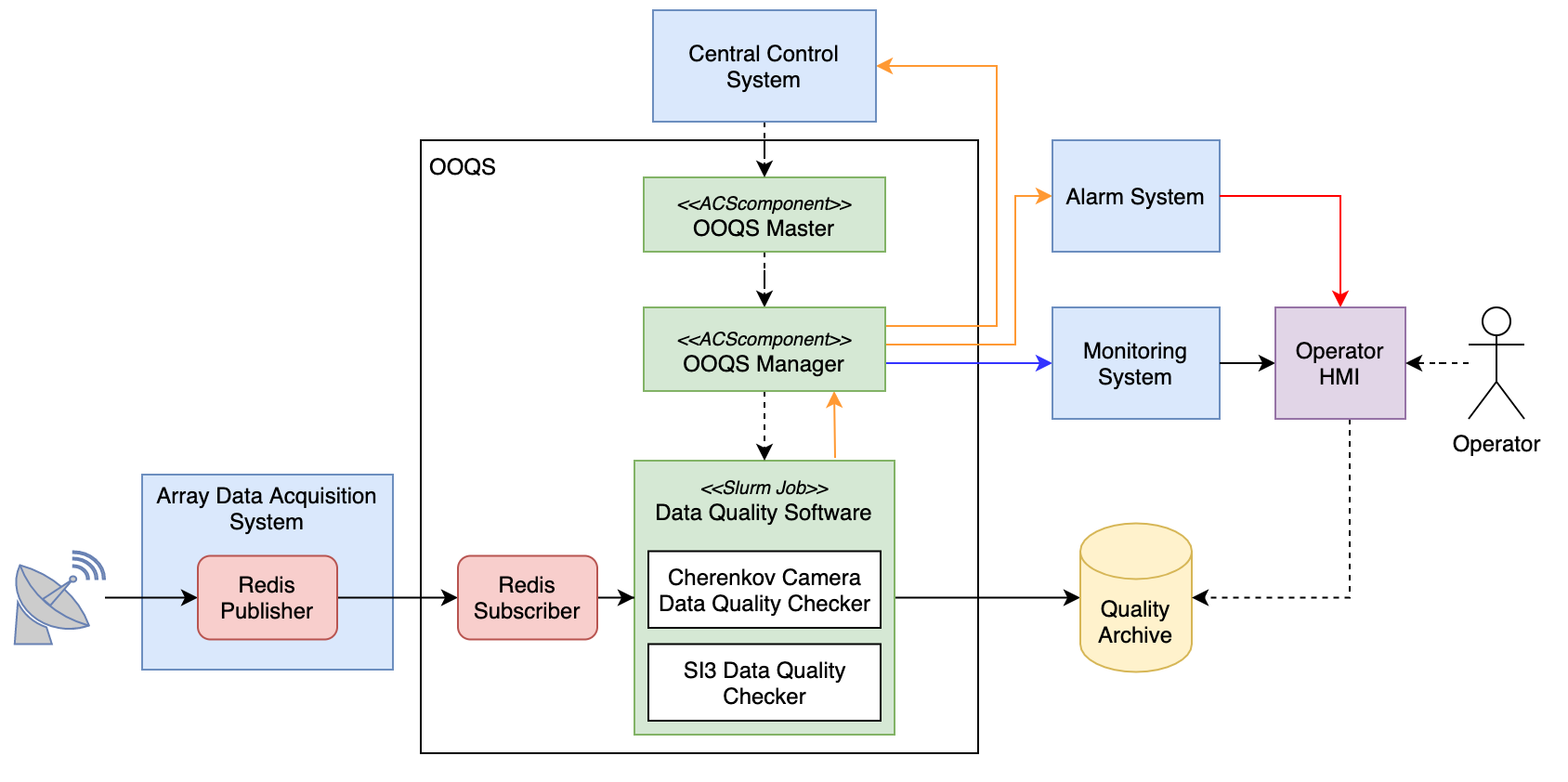}
	\caption{High-level schema of the Online Observation Quality System.}
	\label{fig:scada_arch}
\end{figure*}

Figure \ref{fig:scada_arch} shows a high-level schema of the OOQS architecture and the interfaces with other SCADA  sub-systems. The OOQS is controlled by the Central Control System (part of SCADA) that starts and stops OOQS instances for each telescope through the OOQS Master component. This ACS component is the interface between the Central Control System and the OOQS. It can manage the components' life cycle and monitor them during the operations. The quality checks on the data are performed by the Data Quality Software (CCDQC or SI3DQC) that receives the input data from the ADAS system through the Redis Pub-Sub service (\url{https://redis.io/topics/pubsub}).

Redis is an in-memory key-value database, largely used for its high throughput. The Redis Pub-Sub service implements the Publish-Subscribe paradigm. In this messaging pattern, the senders of messages (publishers) do not send the messages to specific receivers but instead send the messages through channels that can be used by the receivers (subscribers) to filters messages and decide the types of messages to receive. The subscribers do not know the publisher; they subscribe to one or more messages channels. The ADAS implements a Redis Publisher that sends the data acquired by the telescopes to the Redis Subscriber implemented within the OOQS. The ADAS and the OOQS are deployed in different physical servers, and they use this Redis service to share data.

The data quality processes are executed outside the ACS framework. These processes are managed with the Slurm (\url{http://slurm.schedmd.com}) workload manager that can schedule and run multiple processes in parallel to optimise the available resources. Slurm can also be configured in a cluster of servers to execute the jobs in the first available resource shared within the cluster. With this workload manager, it is possible to configure a hierarchy of processes. When a process is completed, its child is started. If during the operation a process crashes, its replacement is started automatically.

Once the OOQS receives the data, the Data Quality Software performs a list of data quality checks. The results of these analyses are saved into the Quality Archive and can be visualised by the Operator through a web Operator Human Machine Interface (HMI). During the quality check analyses, the software can detect abnormal conditions that are sent to the OOQS Manager that forwards them to the Alarm System and to the Central Control System (orange arrows in Figure \ref{fig:scada_arch}). The Central Control System decides how to manage the abnormal conditions by executing automated actions to correct the problem. The Alarm System judges if the abnormal condition requires sending an alarm to the Operator. The Operator can evaluate the anomaly and choose how to proceed. The OOQS Manager is also interfaced with the Monitoring System \cite{2021ICRCCosta} to send information about the software components, such as the processing time or the status of the analyses (blue arrow in Figure \ref{fig:scada_arch}).  

The software used to perform the quality check and all its dependencies (e.g. external packages, services etc.) are installed into a Singularity (\url{https://sylabs.io/singularity/}) container that can be easily deployed in the on-site datacenter. Singularity is an open-source software used for containerisation and developed mainly for scientific high-performance computing to satisfy the needs for reproducibility of the same environment in several hardware and software systems. The containerisation technique has several advantages: (i) the same software with its dependencies can be deployed straightforwardly, (ii) the configurations to obtain the environment are managed with a version control system, and (iii) the container can be easily built using these configurations. The containers can be used to implement a CI workflow as described in Section \ref{sec:dev_plan}.

\subsection{OOQS Master and OOQS Manager}

The OOQS Master is a software component implemented as an ACS Master Component that aims to control the lifecycle of other OOQS ACS components. The OOQS Master is in charge of the startup and the shutdown of all the components and the monitoring of the status of the components during the operation. It can take corrective actions when exceptions are raised. The OOQS Master is interfaced directly with the Central Control System. There are nine OOQS Master components, one for each telescope in the ASTRI MA, that supervise the nine OOQS Manager components.  

The OOQS Manager is an ACS component that controls the software that performs the quality checks on the data acquired by the telescope and sent to the OOQS by the ADAS. The OOQS is composed of nine OOQS Managers, one for each of the nine telescopes. The OOQS Manager is started and controlled by its OOQS Master. The OOQS Manager is interfaced with the Central Control System and the Alarm system to send abnormal conditions detected during the quality checks on input data. It is also interfaced with the Monitoring System to communicate the status of the software that manages (e.g. the number of processes, event rate, processing time etc.). The OOQS Manager can instantiate two types of data quality software: (i) the CCDQC and (ii) the SI3DQC. These two software components are used to perform quality checks on the data acquired during the Cherenkov and intensity interferometry observations, respectively. These two alternative operating modes of the OOQS can not be executed simultaneously, and the OOQS Manager instantiates the appropriate software based on the observation type.  

\subsection{Cherenkov Camera Data Quality Checker}

\begin{figure*}[!htb]
	\centering
	  \includegraphics[width=\textwidth]{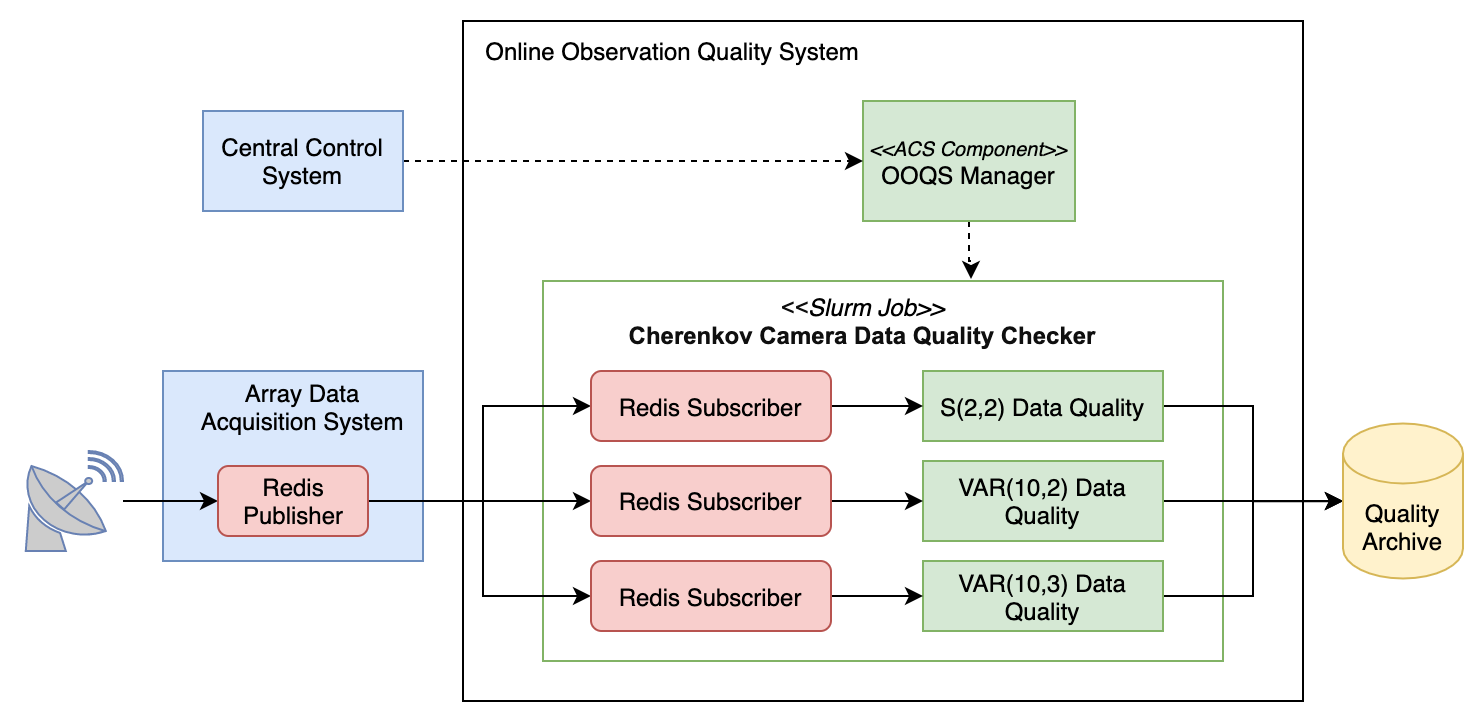}
	\caption{High-level schema of the Cherenkov Camera Data Quality Checker.}
	\label{fig:cherenkov_workflow}
\end{figure*}

Figure \ref{fig:cherenkov_workflow} shows the workflow of the CCDQC. This software component aims to perform a list of quality checks on the data acquired by the Cherenkov cameras and received from the ADAS system. The Central Control System asks the OOQS Manager to activate this software when a new Cherenkov observation is started. This software component submits the analyses to Slurm that manages the workload. The data are shared between the OOQS and the ADAS using the Redis Pub-Sub service described in Sect. \ref{sec:architecture}. The list of data quality checks performed online is defined with the collaboration of the Cherenkov camera designers. These checks aim to obtain fast detection of abnormal conditions that can trigger possible automated or manual corrective actions. The results of the data quality checks are then saved in the Quality Archive to be visualised, through the HMI, by the Operator, which eventually performs further investigations. The inputs for the CCDQC are different data types generated by the Cherenkov camera: the scientific data type, called S(2,2), and two variance data types, the VAR(10,2) and VAR(10,3).

The S(2,2) packet has a size of about 13 kB. It contains the high gain and low gain ADC values for the 64 pixels of the 37 Photon Detection Modules (PDMs) forming each Cherenkov camera, the time of the pixel triggers, temperatures, and other information of the acquisition context. The maximum event rate for each camera is about 600 Hz, and the data rate is about 7.8 MB/s for each telescope or 70.2 MB/s considering the nine telescopes. The VAR have a size of about 9.5 kB, but the event rate is between 1 and 10 Hz; thus, they don't generate a huge data rate. The VAR(10,2) and VAR(10,3) packets are produced by randomly sampling the pixels' values in the absence of a trigger. The variance of these values is calculated separately for the high-gain (VAR(10,2)) and low-gain (VAR(10,3)) chains. The two VAR packets have the same data structure.

The main data quality checks performed on the S(2,2) are (i) calculate the histograms of the trigger number for each camera and each camera's PDM, (ii) check that the pixel values are in a predefined range, (iii) calculate the histogram of the times between two consecutive triggers, (iv) perform the previous analyses with the calibrated data obtained with predefined calibration coefficients, and (v) sample the data to obtain one camera image per second.

The main data quality check performed on the VAR(10,2) and VAR(10,3) are (i) aggregate all the VAR data from the start of the observation, (ii) calculate the ratio between the high-gain and low-gain of each camera PDM, (iii) check if the pointing deviation and the point spread function (PSF) size are inside a nominal range, and (iv) sample the data to obtain one camera image per second. The main purpose of the VAR analysis is to check if it is required a pointing correction, applied in real-time or during the next observation. More details about the Variance methods are described in \cite{2021ICRCIovenitti}.

\subsection{SI3 Data Quality Checker}

\begin{figure*}[!htb]
	\centering
	  \includegraphics[width=\textwidth]{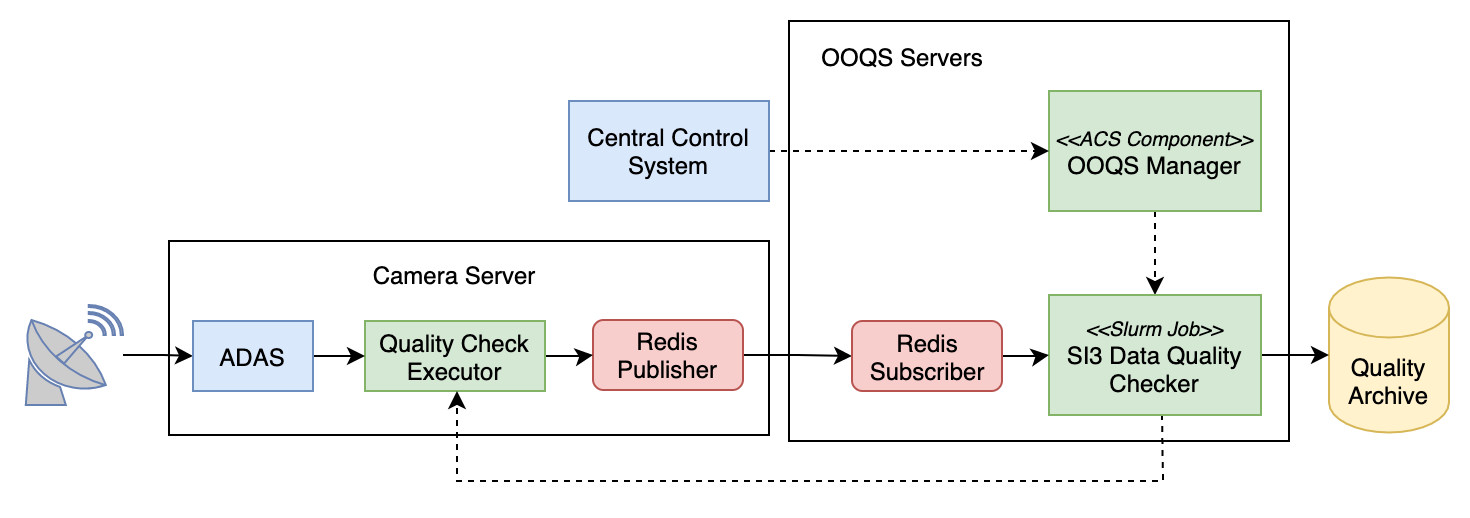}
	\caption{High-level schema of the SI3 Data Quality Checker.}
	\label{fig:si3_workflow}
\end{figure*}

This software component has the goal to perform the online data quality check on the data acquired by the nine SI$^3$ instruments. The event and data rates that this component shall manage are higher than those obtained during the Cherenkov acquisition. Each telescope will have an event rate of 100 Mevents/second and a data rate of 500 MB/s. The event rate for the full array of nine telescopes is about 900 Mevents/second, and the data rate is about 4,5 GB/s. The design of this component (Figure \ref{fig:si3_workflow}) is different from that used for the Cherenkov acquisition because this software shall satisfy these demanding throughput requirements. In the current design, the OOQS Manager controls the SI3 Data Quality Checker that performs the quality checks using a software component called Quality Checker Executor. This software component runs directly into the server where the ADAS system acquires the data (Camera Server) to avoid the data transfer to the servers where the OOQS runs. The OOQS Manager receives back the results of the quality checks performed inside the Camera Server through the Redis Pub-Sub service. In this scenario, the OOQS is a distributed software between different servers. 
  
The main data quality checks performed on the data acquired during the SI$^3$ observations are: (i) extract the event rate for each telescope from the raw data stream, (ii) check the pointing accuracy using the event rate acquired by the four quadrants of each SI$^3$ detector, and (iii) select a random sub-sample of events from the data stream to check the content of the data packet. These checks are not complex from the implementation point of view but shall be performed on a 4,5 GB/s data stream.

\section{Software Development Plan} \label{sec:dev_plan}

The OOQS is developed following the guidelines from the ASTRI-MA Software Development Plan. The software is under version control with the Git software (\url{https://git-scm.com}), and the repository is stored in the GitLab (\url{https://about.gitlab.com}) instance hosted by the INAF ICT department. A CI pipeline is implemented following the DevOps principles \cite{2019arXiv190905409L} using the GitLab CI features to build and test the OOQS software when the source code is updated. The CI pipeline is executed in a container where all the required software and libraries are installed. The execution of the pipelines is performed in a remote server configured as a GitLab Runner (\url{https://docs.gitlab.com/runner/}).

\section{Conclusions}\label{sec:conclusion}

This contribution describes the design and the software architecture of the OOQS system, part of the SCADA system, for the ASTRI MA. This software aims to perform data quality checks in real-time during the data acquisition from the nine telescopes of the ASTRI MA. The OOQS analyses data during two observing modes: the Cherenkov observing mode and the stellar intensity interferometry observing mode. These observing modes are alternative and can not be executed at the same time. For this reason, the OOQS instantiates the software specifically designed for the data quality checks of the data acquired by the Cherenkov cameras or by the SI$^3$ instruments. The data rates of the two observing modes are different, and in the case of the SI$^3$ observations, it can reach a maximum of 4.5 GB/s for the whole array. 

The software architecture of the OOQS is designed to satisfy the performance requirements while executing the data quality checks during the ASTRI MA observations. This system is critical because it can detect abnormal conditions and trigger external systems to perform automated corrective actions. The Operator can visualise the OOQS results through the HMI during the supervision of the observations and take corrective actions if needed.

The software and its dependencies are installed into a Singularity container. The containerisation is essential for implementing the CI pipeline and deploying the OOQS in different hardware configurations exploiting the DevOps techniques.

\acknowledgments

This work was conducted in the context of the ASTRI Project. This work is supported by the Italian Ministry of University and Research (MUR) with funds specifically assigned to the Italian National Institute for Astrophysics (INAF). We acknowledge support from the Brazilian Funding Agency FAPESP (Grant 2013/10559-5) and from the South African Department of Science and Technology through Funding Agreement 0227/2014 for the South African Gamma-Ray Astronomy Programme. This work has been supported by H2020-ASTERICS, a project funded by the European Commission Framework Programme Horizon 2020 Research and Innovation action under grant agreement n. 653477. IAC is supported by the Spanish Ministry of Science and Innovation (MICIU).



\bibliographystyle{JHEP}
\bibliography{paper}

%
%
%

\end{document}